\documentclass[11pt,times,letter]{article}

\usepackage[margin=1in]{geometry} 
\usepackage{parskip}
\usepackage{longtable,color,natbib}
\usepackage{booktabs} 
\usepackage{array} 
\usepackage{paralist} 
\usepackage{verbatim} 
\usepackage{subfigure} 
\usepackage{appendix}
\usepackage{amssymb}
\usepackage{amsthm}
\usepackage{amsmath}
\usepackage{stmaryrd}

\definecolor{linkcolor}{rgb}{0.1,0,0.7}
\definecolor{urlcolor}{rgb}{1,0,0}
\usepackage[unicode=true,colorlinks,citecolor=linkcolor,linkcolor=linkcolor,urlcolor=blue]{hyperref}
\usepackage{graphicx} 
\usepackage{epstopdf}
\usepackage{bm}

\def\esssup_#1{\underset{#1}{\mathrm{ess\,sup\, }}}
\def\essinf_#1{\underset{#1}{\mathrm{ess\,inf\, }}}
\def\argmax_#1{\underset{#1}{\mathrm{arg\,max\, }}}
\def\argmin_#1{\underset{#1}{\mathrm{arg\,min\, }}}

\newcommand{\cadlag}{c\`{a}dl\`{a}g }

\newtheorem{theorem}{Theorem}
\newtheorem{definition}{Definition}

\allowdisplaybreaks[4]

\title{Erratum to ``On the market viability under proportional transaction costs"}

\author{Erhan Bayraktar \thanks{Email: erhan@umich.edu, Department of Mathematics, University of Michigan, 530 Church Street, Ann Arbor, MI48109, USA.}
\and
Xiang Yu \thanks{Email: xiang.yu@polyu.edu.hk, Department of Applied Mathematics, The Hong Kong Polytechnic University, Hung Hom, Kowloon, Hong Kong.}
}
\date{\vspace{-0.5cm}}

\begin{document}

\maketitle

\begin{abstract}
This short note aims to point out mistakes in one of the implications for Theorem $2.8$ in Bayraktar and Yu [Mathematical Finance, 28 (2018), pp. 800-838], which weakens the statement of this theorem.
\end{abstract}

\vspace{0.2in}

Consistent price systems (CPS), introduced by \cite{JK1995} and \cite{CK1996}, play the role of the dual elements in the fundamental theorem of asset pricing with proportional transaction costs. Sufficient conditions to guarantee the existence of a CPS have been extensively studied in the literature, which are closely related to the stringent no arbitrage condition such as No Free Lunch with Vanishing Risk (NFLVR), see among \cite{Guasoni08}, \cite{BS2010}, \cite{RSch2010},
\cite{MMS2011}, \cite{SV2011}, \cite{Guasoni12}. On the other hand, in the frictionless market models, some weaker model assumptions such as No Unbounded Profit with Bounded Risk (NUPBR) have been studied when the market viability can still be verified in the sense that the classical option hedging and utility maximization problems can be solved, see for instance \cite{KK2007}, \cite{Bech2001}, \cite{CL2007}, \cite{IP2015}, and et al.  

In our paper \cite{BY2018}, motivated by the existing studies on minimal market conditions in frictionless market models, the goal is to provide and verify some weaker market conditions such that the market viability can hold in the model with proportional transaction costs without requiring the existence of (strictly) consistent price systems. In particular, we propose to consider the dual element called Strictly Consistent Local Martingale Systems (SCLMS) $(\tilde{S}, Z)$ in the following definition. 

\begin{definition}\label{SCLMS}
Given the stock price $(S_t)_{t\in[0,T]}$ with transaction cost $\lambda_t\in (0.1)$ a.s. for all $t\in[0,T]$, we call a pair $(\tilde{S}, Z)$ a consistent local martingale system (CLMS) if $\tilde{S}$ is a semimartingale satisfying
\begin{equation}
(1-\lambda_t)S_t \leq \tilde{S}_t\leq (1+\lambda_t)S_t,\ \ \mathbb{P}\text{-a.s.},\ \ \forall t\in[0,T],\nonumber
\end{equation}
and there exists a strictly positive local martingale $Z_t$ with $Z_0=1$ such that $\tilde{S}_tZ_t$ is a local martingale. We shall denote $\mathcal{Z}_{\text{loc}}(\lambda)$ the set of all CLMS with transaction cost $(\lambda_t)_{t\in[0,T]}$. Moreover, if
\begin{equation}
\inf_{t\in[0,T]} \left(\lambda_tS_t-|S_t-\tilde{S}_t|\right)>0,\ \ \mathbb{P}\text{-a.s.}, \nonumber
\end{equation}
we shall call the pair $(\tilde{S}, Z)$ a SCLMS and denote by $\mathcal{Z}_{\text{loc}}^{s}(\lambda)$ the set of all SCLMS.
\end{definition}

In our main result Theorem 2.8 in \cite{BY2018}, we plan to establish the equivalence between the existence of SCLMS and some weaker market conditions, namely the NUPBR condition and the No Local Arbitrage with Bounded Portfolios (NLABP) condition with the transaction cost $\lambda$ in the robust sense of Definition 2.7 in \cite{BY2018}. In particular, the definition of NLABP is given in \cite{BY2018} as below.

\begin{definition}\label{DefnNLA}
We say that $S$ satisfies No Local Arbitrage with Bounded Portfolios (NLABP) with the transaction cost $\lambda$ if there exists a sequence of stopping times $\tau_n\nearrow T$ as $n\rightarrow\infty$ such that for each $n\in\mathbb{N}$, we can not find $(\phi^{0,n}, \phi^{1,n})\in\mathcal{A}^{\text{bd}}(\lambda)$ which satisfies
\begin{equation}\label{localabpcon}
\mathbb{P}\left(V_{\tau_n}^{\text{liq}, 0}(\phi^{0,n},\phi^{1,n})\geq 0\right)=1\ \ \ \text{and}\ \ \ \mathbb{P}\left(V_{\tau_n}^{\text{liq}, 0}(\phi^{0, n}, \phi^{1,n})>0\right)>0,
\end{equation}
where we denote
\begin{equation}
\mathcal{A}_x^{\text{bd}}(\lambda) \triangleq \{(\phi^0, \phi^1): |\phi_t^1|\leq M,\ \mathbb{P}\text{-a.s.},\ t\in[0,T]\ \text{for some $M>0$ where}\ (\phi^0, \phi^1)\in\mathcal{A}_x(\lambda)\},
\end{equation}
and $\mathcal{A}^{\text{bd}}=\bigcup_{x\geq 0}\mathcal{A}_x^{\text{bd}}$.
\end{definition}

The next result is the main theorem in \cite{BY2018}.

\begin{theorem}[Theorem $2.8$ in \cite{BY2018}]\label{mainthm}
The following two assertions are equivalent.
\begin{itemize}
\item[(1)] $S$ satisfies the NUPBR and NLABP conditions with the transaction cost $\lambda$ in the robust sense.
\item[(2)] There exists a SCLMS $(\tilde{S},Z)$ for the market with transaction cost $\lambda$, i.e., $\mathcal{Z}^s_{\text{loc}}(\lambda)\neq\emptyset$.
\end{itemize}
\end{theorem}

In the definition of the NLABP condition, we require the portfolio process $ |\phi_t^1|\leq M$ uniformly in time for some constant $M>0$ in order to facilitate some technical proofs. In particular, in the proof of the direction  $(1)\Rightarrow (2)$, we have used some arguments based on $ |\phi_t^1|\leq M$ leading to the contradiction to NLABP condition. Unfortunately, this constant bound $M$ in our proposed NLABP condition also becomes an obstacle in some of other arguments in the direction $(1)\Rightarrow (2)$, which we will elaborate on.

In showing the direction of $(1)\Rightarrow (2)$, we note that if the stock price process $S$ satisfies the NUPBR and NLABP conditions with the transaction cost $\lambda$ in the robust sense, Lemma $3.9$ in \cite{BY2018} first gives the existence of the auxiliary pair $(\breve{S}, \breve{\lambda})$ such that $\breve{S}$ satisfies the RNUPBR condition and NLABP condition with the transaction cost $\breve{\lambda}$. We then plan to work with the pair $(\breve{S}, \breve{\lambda})$. 

As $\breve{S}$ is locally bounded because the stock price $(S_t)_{t\in[0,T]}$ is locally bounded, there exist an increasing sequence of constants $\alpha(n)\nearrow+\infty$ and an increasing sequence of stopping times $\rho_n\nearrow T$ such that $\breve{S}_t\leq \alpha(n)$ on the stochastic interval $\llbracket 0,\rho^n\rrbracket$. To align with the definition of NLABP condition in Definition $\ref{DefnNLA}$, we first consider a sequence of stopping times $\{\tau_n\}_{n\in\mathbb{N}}$ in Definition $\ref{DefnNLA}$, and then define the new sequence of stopping times
\begin{equation}\label{goodseqtau}
\bar{\tau}_0\triangleq 0,\ \ \ \bar{\tau}_n\triangleq \tau_n\wedge \rho_n, \quad \text{for $n\in\mathbb{N}$.}
\end{equation}
As a result, the process $\breve{S}$ is bounded up to $\bar{\tau}_n$ for each $n\in\mathbb{N}$.

We then consider the set $\mathbf{C}_M^{\bar{\tau}_n}(\breve{S},\breve{\lambda})$ for some fixed large level $M>0$ on $\llbracket 0,\bar{\tau}_n\rrbracket$, where we define
\begin{align}\label{boundMCset}
\mathbf{C}_M^{\bar{\tau}_n}(x; \breve{S}, \breve{\lambda})\triangleq \{ &V: V\leq V_{\bar{\tau}_n}^{\text{liq}, 0}(\phi^0, \phi^1; \breve{S}, \breve{\lambda})\ \text{where}\ (\phi^0,\phi^1)\in\mathcal{A}_x(\breve{S}, \breve{\lambda})\\
& \text{and}\ |\phi_t^1|\leq M,\ \mathbb{P}\text{-a.s. on}\ \llbracket 0,\bar{\tau}_n\rrbracket \},
\end{align}
and $\mathbf{C}_M^{\bar{\tau}_n}(\breve{S}, \breve{\lambda}) \triangleq \bigcup_{x\geq 0}\mathbf{C}_M^{\bar{\tau}_n}(x; \breve{S}, \breve{\lambda})$.

By Lemma $3.8$ in \cite{BY2018}, we have that $\mathbf{C}_M^{\bar{\tau}_n}(\breve{S}, \breve{\lambda})\cap\mathbb{L}^{\infty} $ is Fatou closed; and from Lemma $3.6$ in \cite{BY2018}, it follows that $\mathbf{C}_M^{\bar{\tau}_n}(\breve{S},\breve{\lambda})\cap\mathbb{L}_+^{\infty}=\{0\}$. A key step in the proof of this direction is the claim that we can apply Lemma $5.5.2$ in \cite{KSaf2009} (which relates Fatou-closedness to weak-$*$ closedness) and Kreps-Yan separation theorem (see, e.g., Theorem $B.3$ in \cite{Guasoni12}) to conclude the existence of a probability measure $\mathbb{Q}^n$ equivalent to $\mathbb{P}$ such that for any $V\in\mathbf{C}_M^{\bar{\tau}_n}(\breve{S},\breve{\lambda}) \cap \mathbb{L}^{\infty}$, we have $\mathbb{E}^{\mathbb{Q}^n}[V]\leq 0$ on page $815$ of \cite{BY2018}.

We appreciate Prof. Martin Schweizer and his student for pointing out that our previous argument to conclude the existence of probability measure $\mathbb{Q}^n$ is incorrect because our constructed set $\mathbf{C}_M^{\bar{\tau}_n}(\breve{S},\breve{\lambda}) \cap \mathbb{L}^{\infty}$ is not a convex cone, which is a key assumption to apply the Kreps-Yan separation theorem. We unfortunately cannot fix this error at the moment which is why we are submitting this erratum. As a result, our proposed NLABP condition involving the constant bound $M$ in Definition \ref{DefnNLA} may not be a reasonably weak NA condition for the existence of SCLMS.  

In addition to the previous mistake, we also discovered an additional mistake  in the same proof of the direction $(1)\Rightarrow (2)$ on page $816$ of \cite{BY2018}. Therein, we plan to construct some predictable, non-positive, self-financing portfolio processes $(\tilde{\phi}^0, \tilde{\phi}^1)$ such that $V^{\text{liq}, 0}_{\bar{\tau}_n}(\tilde{\phi}^0,\tilde{\phi}^1)\in\bar{\mathbf{C}}_M^{\bar{\tau}_n}(\breve{S},\breve{\lambda})\cap\mathbb{L}^{\infty}$ and $\mathbb{E}^{\mathbb{Q}^n}[V^{\text{liq}, 0}_{\bar{\tau}_n}(\tilde{\phi}^0, \tilde{\phi}^1)]\leq 0$ in order to conclude $\mathbb{E}^{\mathbb{Q}^n}[\breve{S}_{\sigma}(1+\breve{\lambda}_{\sigma})|\mathcal{F}_{\eta}]\geq \breve{S}_{\eta}(1-\breve{\lambda}_{\eta})$. By Lemma $3.10$ in \cite{BY2018}, we obtain the existence of a \cadlag martingale $\tilde{S}^n$ under $\mathbb{Q}^n$ such that
\begin{equation}\label{existSn}
\breve{S}_t(1-\breve{\lambda}_t)\leq \tilde{S}_t^n\leq \breve{S}_t(1+\breve{\lambda}_t),\ \ \text{a.s. for}\ \ 0\leq t\leq\bar{\tau}_n.
\end{equation}
Our plan is to paste the processes $\{\tilde{S}^n\}_{n\in\mathbb{N}}$ over the whole interval $[0,T]$ in a proper way to obtain the SCLMS.

However, the constructed portfolio $(\tilde{\phi}^0, \tilde{\phi}^1)$ (see also equation $(3.17)$ on page $816$ of \cite{BY2018}) 
\begin{equation}\label{specialphi11}
\begin{split}
\tilde{\phi}^0&\triangleq \left[(1-\breve{\lambda}_{\eta})\breve{S}_{\eta}\mathbf{1}_{\rrbracket \eta, \sigma\llbracket}(t)+\Big((1-\breve{\lambda}_{\eta})\breve{S}_{\eta}-(1+\breve{\lambda}_{\sigma})\breve{S}_{\sigma}\Big)\mathbf{1}_{\llbracket \sigma, \bar{\tau}_n\rrbracket}\right]\mathbf{1}_A,\\
\tilde{\phi}^1&\triangleq -\mathbf{1}_{\rrbracket \eta, \sigma\llbracket}(t)\mathbf{1}_A,
\end{split}
\end{equation}
for some stopping time $\eta\leq \sigma\leq \bar{\tau}_n$ and $A\in\mathcal{F}_{\eta}$ is in fact not predictable. The same mistake holds for the constructed portfolio process $(\hat{\phi}^0, \hat{\phi}^1)$ on the same page. Therefore, we have to construct or consider some other predictable admissible portfolio processes to continue the proof therein. 

In conclusion, due to these two mistakes, the implication $(1)\Rightarrow (2)$ of Theorem $2.8$ in \cite{BY2018} does not work. Or our proposed NLABP condition likely is not weak enough to guarantee the existence of SCLMS.

\ \\
\ \\

\end{document}